\documentclass[11pt]{amsart}
\theoremstyle{plain}
\usepackage{pgf}
\usepackage{amsmath,amssymb,amscd,cite}
\usepackage{graphicx}
\usepackage{xcolor}
\usepackage{enumerate}
\newtheorem{thm}{Theorem}[section]
\newtheorem{prop}[thm]{Proposition}
\newtheorem{lem}[thm]{Lemma}
\newtheorem{cor}[thm]{Corollary}
\newtheorem{defi}[thm]{Definition}
\newcommand{\pf}{{\bf Proof. \ }}

\newtheorem{rem}[thm]{Remark}
\newtheorem{ex}[thm]{Example}

\begin{document}
\author{Nabil Bennenni, Kenza Guenda and Sihem Mesnager }

\title{New DNA Cyclic Codes over Rings}

\maketitle

\begin{abstract}
This paper is dealing with DNA cyclic codes which play an important role in DNA computing and have attracted a particular  attention in the literature.\\
Firstly, we introduce a new family of DNA cyclic codes over the ring $R=\mathbb{F}_2[u]/(u^6)$. Such codes have  theoretical advantages as well as several applications in DNA computing. A direct link between the elements of such a ring and the $64$ codons used in the amino acids of the living organisms is established. Such a correspondence allows us to extend the notion of the edit distance to the ring $R$ which is useful for the correction of the insertion, deletion and substitution errors. Next, we define the Lee weight, the Gray map over the ring $R$ as well as the binary image of the cyclic DNA codes allowing the transfer of studying DNA codes into studying binary codes. Secondly, we introduce another new family of DNA skew cyclic codes constructed  over the ring $\tilde {R}=\mathbb{F}_2+v\mathbb{F}_2=\{0,1,v,v+1\}$ where $v^2=v$ and study their property of being reverse-complement. We show that the obtained code is derived from the cyclic reverse-complement  code over the ring $\tilde {R}$. We shall provide the binary images and present some explicit examples of such codes.
\end{abstract}
\section{Introduction}
Deoxyribonucleic acid (DNA) contains the genetic program for the biological development of life. DNA is formed by strands linked together and twisted in the shape of a double helix. Each strand is a sequence of four possible nucleotides, two purines; adenine $(A)$, guanine $(G)$ and two pyrimidines; thymine $(T)$ and
cytosine $(C)$. The ends of the DNA strand are chemically polar with $5'$ and $3'$ ends. Hybridization, known as base pairing, occurs when a strand binds to another strand, forming a double strand of DNA. The strands are linked following the Watson-Crick model;
every $(A)$ is linked with a $(T)$, and every $(C)$ with a $(G)$, and vice versa. We denote by $\hat{x}$ the complement of $x$ defined as follows, $\hat{A}=T,\hat{T}=A,\hat{G}=C$ and $\hat{C}=G$ (for instance  if $x=(AGCTAC)$, then its complement $\hat{x}=(TCGATG)$.
\newline
DNA computing combines genetic data analysis with science of the computation in order to tackle computationally difficult problems. This new field started by Leonard Adleman \cite{A3}, who solved a hard computational problem by DNA molecules in a test tube.
Several authors have contributed to provide constructions of cyclic DNA codes over fixed rings.
 In \cite{A02,A4}, the authors gave DNA cyclic codes over finite field with four elements. Further, Siap et al. have studied in  \cite{Siap} cyclic DNA codes over the ring $\mathbb{F}_2[u]/(u^2-1)$ using the deletion distance. More recently, Guenda et al. have studied in \cite{GGS} cyclic DNA codes of arbitrary length over the ring $\mathbb{F}_2[u]/(u^4-1)$. Those codes have several applications as well as high hybridization energy.
In this paper we  firstly consider the DNA codes of length $n$ over the ring $R=\mathbb{F}_2[u]/(u^6)$. The ring $R$ is a principal commutative ring with $64$ elements. With four possible bases, the three nucleotides can give $4^3=64$ different possibilities, called codons. These combinations are used to specify the $20$ different amino acids used in the living organisms \cite{gene2}. To this end, we construct a one-to-one correspondence between the elements of $R$ and the 64 codons over the alphabet $\{A,G,C,T\}^3.$ Such a correspondence allows us to extend the notion of the edit distance to the ring $R$.
The edit distance is an important combinatorial notion for the DNA strands. It can be used for the correction of the insertion, deletion and substitution errors between the codewords. This it is not the case for the Hamming, deletion, and the additive stem distances. For that in this paper, we design  cyclic reverse-complement DNA codes over the ring $R$ with designed edit distance $D$. We also give some upper and lower bounds on $D$. We define a Lee weight and a Gray map over $R$. The images of our DNA codes under the mapping are quasi-cyclic codes of  index $6$ and of  length $6n$ over the alphabet $\{A,G,C,T\}$.
 There are several advantages in using codes over the ring $R$. We list some of them below:
\begin{enumerate}
\item There exists a one-to-one correspondence between the codons and the elements of the ring $R$.
\item A code over $R$ can contains more codewords than codes of the same length over fields.
\item The factorization of $x^n-1$ is the same over the field $ \mathbb{F}_2$ but  is not the same over other rings. This fact simplifies the construction of cyclic codes over $R$.
\item The structure of the cyclic codes of any length over $R$ is well-known ~\cite{GGS1}, whereas little is know concerning the structure of cyclic codes of any length over rings.
\item The cyclic character of the DNA strands is desired because the genetic code should represent an equilibrium status \cite{gene}. Another advantage of cyclic codes, as indicated by Milenkovic and Kashyap~\cite{ImageB}, is that the complexity of the dynamic programming algorithm for testing DNA codes for secondary structure will be less  for cyclic codes.
\item The binary image of the cyclic codes over $R$ under our Gray map are linear quasi-cyclic codes.

\end{enumerate}
In the second part of the paper, we study the  skew cyclic DNA codes over the ring $\tilde R=\mathbb{F}_2+v\mathbb{F}_2=\{0,1,v,v+1\}$, where $v^2=v.$ The codes obtained satisfy the reverse-complement. Further we give the binary images of the skew cyclic DNA codes and provide some examples. The advantage of  studying the reversible DNA code in  skew polynomial rings is to exhibit several factorizations. Therefore, many reverse-complement DNA code could be obtained  in a skew polynomial ring (which is not the case in a commutative ring).\\

This paper is organized as follows. In Section 2 we start by presenting some preliminaries results as well as the one-to-one correspondence between the element of the ring $R=\mathbb{F}_2[u]/(u^6)$ and the codons.
Next, we give the algebraic structure of the cyclic codes over $R=\mathbb{F}_2[u]/(u^6)$ and we study the DNA cyclic codes and reverse-complement of these codes. Moreover, we define the Lee weight related to such codes and give the binary image of the cyclic DNA code. Some explicit examples of such codes are presented.
 In Section 3, we describe a skew cyclic DNA codes over $\tilde R=\mathbb{F}_2+v\mathbb{F}_2=\{0,1,v,v+1\}$ where $v^2=v$. We study their property of being reverse-complement and provide explicit examples of such codes with minimum Hamming distance.

\section{Cyclic DNA Codes over $R=\mathbb{F}_2[u]/(u^6)$}
\subsection{Some Background and Preliminaries}
 The ring considered in this section is
\newline
 $$R=\mathbb{F}_2[u]/(u^6)=\{a_0+a_1u+a_2u^2+a_3u^3+a_4u^4+a_5u^5; a_i\in\mathbb{F}_2,u^6=0\}.$$
 It is a commutative ring with 64 elements. It is a principal local ideal ring with maximal ideal $\langle u\rangle.$ The ideals of $R$ satisfy the following inclusions
 $$\langle0\rangle=\langle u^6\rangle\subsetneq\langle u^5\rangle\subsetneq\langle u^4\rangle\subsetneq\langle u^3\rangle\subsetneq\langle u^2\rangle\subsetneq\langle u\rangle\subsetneq\langle R\rangle.$$

 Since the ring $R$ is of the cardinality 64, then we can construct a one-to-one correspondence between the elements of $R$ and the 64 codons over the alphabet $\{A,G,C,T\}^3$ by the map $\phi$,  this is given in Table 1. A simple
verification give that for all $x\in R,$ we have
 \begin{equation}
 \label{eq:hat}
 x+\hat{x}=u^5+u^4+u^3+u^2+u+1.
 \end{equation}
 \begin{table}[h]
\caption{ Identifying Codons with the Elements of the Ring $R.$}
\begin{center}
\tiny{
\begin{tabular}{cccccccc}
  \hline

  CCC & $u^5+u^4+u^3+u^2+u+1$ & GGG & 0      & ACT & $u^3+u^2+1$ & GTC & $u^4+u^2+u+1$          \\
  GGA & $u^5+u^4+u^3+u^2+u$   & CCT & 1      & ACG & $u^3+u^2+u$ & ACA & $u^3+u^2+u+1$         \\
  GGC & $u^5+u^4+u^3+u^2+1$   & CCG & $u$    & TTT & $u^4+u^2+1$ & GAC & $u^5+u^3+u^2+1$          \\
  GGT & $u^5+u^4+u^3+u^2$     & CCA & $u+1$  & TTG & $u^4+u^2+u$ & AGG & $u^5+u^3+u+1$          \\
  AGG & $u^5+u^4+u^3+u+1$     &TCC & $u^2$   & CTA & $u^4+u+1$   & GAT & $u^5+u^3+u^2$        \\
  CGG & $u^5+u^4+u^2+u+1$     & GCC & $u^3$  & GTT & $u^4+u^3+1$ & GTA & $u^4+u^3+u+1$            \\
  GAG & $u^5+u^3+u^2+u+1$     & CTC & $u^4$  & GTG & $u^4+u^3+u$ & ATT & $u^4+u^3+u^2+1$            \\
  AGA & $u^5+u^4+u^3+u$       & TCT & $u^2+1$& TCA & $u^2+u+1$   & ATA & $u^4+u^3+u^2+u$            \\
  AGC & $u^5+u^4+u^3+1$       & TCG & $u^2+u$& CAA & $u^5+u^2+u$ & ATC & $u^4+u^3+u^2$            \\
  ATG & $u^4+u^3+u^2+u+1$     &TAC & $u^5$   & CAC & $u^5+u^2+u$ & TGA & $u^5+u^4+u$             \\
  AGT & $u^5+u^4+u^3$         &TAT & $u^5+1$ & GCA & $u^3+u+1$   & AAT & $u^5+u^2+u+1$           \\
  CGA & $u^5+u^4+u^2+u$       & GCT & $u^3+1$& TTA & $u^4+u^3$   & AAA & $u^5+u^3+u$             \\
  CGC & $u^5+u^4+u^2+1$       & GCG & $u^3+u$& ACC & $u^3+u^2$   & TGC & $u^5+u^4+1$             \\
  CGT & $u^5+u^4+u^2$         & TAA & $u^5+u$& CAT & $u^5+u^2$   & AAC & $u^5+u^3+1$               \\
  TGG & $u^5+u^4+u+1$         & CTG & $u^4+u$& TGT & $u^5+u^4$   & TCC & $u^4+u^2$               \\
  GAA & $u^5+u^3+u^2+u$       & CTT & $u^4+1$& CAG & $u^5+u^3$   & TAG & $u^5+u+1$              \\
  \hline
\end{tabular}
}
\end{center}
\end{table}
Now, since $R^n$ is an $R$-module, a linear code over $R$ of length $n$ is a submodule $\mathcal{C}$ of $R^n$.
An $(n,k)$ linear block code of dimensions $n = ml$, is called \textbf{quasi-cyclic} if every cyclic shift of a codeword by $l$ symbol yields another codeword.
Define the Hamming weight of the codeword $x\in\mathcal{C}$ as $w_H(x)=n_{a_i}(x),$ where $a_i\in R^*.$ The Hamming distance $d_H(x,y)$ between the vector $x$ and $y$ is $w_H(x-y).$
Let $x=x_0x_1\ldots x_{n-1}$ be a vector in $R^n$. The reverse of $x$ is defined as $x^r=x_{n-1}x_{n-2}\ldots x_{1}x_0$, the complement of $x$ is $x^c=\hat{x}_0\hat{x}_1\ldots \hat{x}_{n-1}$, and the reverse-complement, also called the Watson-Crick complement (WCC) is defined as $x^{rc}=\hat{x}_{n-1}\hat{x}_{n-2}\ldots \hat{x}_{1} \hat{x}_{0}$. A code
$\mathcal{C}$ is said to be \textbf{reversible} if for any $x\in \mathcal{C}$  we have $x^{r}\in \mathcal{C}$.
$\mathcal{C}$ is said to be \textbf{reverse-complement} if for any $x\in \mathcal{C}$ we have $x^{rc} \in \mathcal{C}$.\\

We shall use  the {\em edit distance} for correction of the insertion, deletion and substitution errors in codewords. It is the minimum number of the operations (insertion, substitution and deletion) required to transform one string into the other.  The edit distance can be defined as follows ~\cite{edit}.

Let $\mathcal{A}$ and $\mathcal{B}$ be finite sets of distinct symbols
and let $x^t\in \mathcal{A}^t$ denotes an arbitrary string of length $t$ over $\mathcal{A}$. Then $x_i^j$ denotes the substring of $x^t$ that begins at position $i$ and ends at position $j.$
The edit distance is characterized by a triple
$\langle\mathcal{A},\mathcal{B},c\rangle$ consisting of the finite sets
$\mathcal{A}$ and $\mathcal{B}$, and the primitive function
$c:E\rightarrow \mathbb{R}_+,$ where $\mathbb{R}_+$ is the set of nonnegative reals,
$E=E_s\cup E_d \cup E_i$ is the set of primitive edit operations,
$E_s=\mathcal{A}*\mathcal{B}$ is the set of substitutions,
$E_d=\mathcal{A}*{E}$ is the set of deletion and
$E_i=E\times\mathcal{B}$ is the set of insertions.
Each triple $\langle\mathcal{A},\mathcal{B},c\rangle$ induces a distance
function $d_c:\mathcal{A}^{*}\times\mathcal{B}^*\rightarrow
\mathbb{R}_+$ which  maps a string $x^t$ to a nonnegative value, defined as follows.
\begin{defi}
The edit distance $d_c(x^t,y^v)$ between two strings
$x^t\in \mathcal{A}^t$ and $y^v\in \mathcal{B}^v$ is
defined recursively as
\[
d_c(x^t,y^v)=\min\left\{
                   \begin{array}{l}
                     c(x_t,y_v)+d_c(x^{t-1},y^{v-1}); \\
                     c(x_t,\epsilon)+d_c(x^{t-1},y^v); \\
                     c(\epsilon,y_v)+d_c(x^t,y^{v-1});
                   \end{array}
                 \right.
\]
where $d_c(\epsilon,\epsilon)=0$ if $\epsilon$ denotes the empty string of length $n$.
\end{defi}
It is easy to check the following bounds on the edit distance $d_c$.
\begin{prop}
\label{prop:edit}
Assume that $X$ and $Y$ are strings in $R^n$. Then the following holds:
\begin{itemize}
\item[(i)] $d_c(\phi(X),\phi(Y))\leq n$;
\item[(ii)] $d_c(\phi(X),\phi(Y))\leq d_H(\phi(Y),\phi(X));$
\item[(iii)] $d_c(\phi(X),\phi(\hat{Y}))=d_c(\phi(Y),\phi(\hat{X})). $
\end{itemize}
\end{prop}
\subsection{Cyclic codes over $R=\mathbb{F}_2[u]/(u^6)$}
In this subsection we give the algebraic structure of the cyclic code of arbitrary length over the ring $R$. We start by giving the definition of cyclic code over this ring.
\newline
Let $\mathcal{C}$ be a code over $R$ of the length $n$. A codeword $\mathcal{C}=(c_0,c_1,\cdot\cdot\cdot,c_{n-1})$ of $\mathcal{C}$  is viewed as a polynomial $c_0+c_1x+\cdot\cdot\cdot+c_{n-1}x^{n-1}\in R[x]$. Let $\tau$  be the cyclic shift acting on the codewords of $\mathcal{C}$ in the following way: $$\tau(c_0,c_1,\cdot\cdot\cdot,c_{n-1})=(c_{n-1},c_0,c_1,\cdot,\cdot,\cdot,c_{n-2}).$$
Recall that linear code $\mathcal{C}$ is cyclic if $\mathcal{C}$ is invariant under permutation $\tau:c(x)\mapsto xc(x)$ modulo ($x^n-1$).

The following result gives the structure of the cyclic code of arbitrary length.

\begin{thm}\label{thm:struct2}(\cite{Dinh,GGS1})
Let $\mathcal{C}$  be a cyclic code of  arbitrary length  $n$ over the ring $R$.
\newline
$(i)$ If $n$ is odd. Then there exist six polynomials $f_0,f_1,f_2,f_3,f_4,f_5$ such that  $f_5|f_4|f_3|f_2|f_1|f_0|x^n-1$ and $\mathcal{C}=\langle f_0,uf_1,u^2f_2,u^3f_3,u^4f_4,u^5f_5\rangle;$
\newline
$(ii)$ If $n=m2^s$ such that $gcd(m,p)=1$. Then the cyclic codes of  length  $n=m2^s$ over $R$ are the ideals generated by $\langle f_0,uf_1,u^2f_2,u^3f_3,u^4f_4,u^5f_5\rangle$, where $f_i|f_0$ and $f_0$ is divisor of $x^n-1$ in $\mathbb{F}_2$.
\end{thm}

Let denote by $K$ the field $R/(u)$. We have the following canonical ring morphism
\begin{center}
 $-:R[x]\rightarrow K[x]$; $f\mapsto \overline{f}=f \pmod u$
 \end{center}
The rank of $\mathcal{C}$ is defined as
$$ k(\mathcal{C})=\sum_{i=0}^5k_i,$$
where the $k_i$ are such that $|C|=|K|^{\sum_{i=0}^5 (5-i)k_i}$.
The submodule quotient of $\mathcal{C}$ by $v\in R$ is the code
 $$ (\mathcal{C}:u^i)=\{v\in R^n|u^iv\in\mathcal{C}\}.$$
 Thus we have the following tower of linear codes over $R$.
\begin{equation}
\label{eq:eq2}
\mathcal{C}=(\mathcal{C}:u)\subseteq (\mathcal{C}:u^2)\subseteq (\mathcal{C}:u^3)\subseteq (\mathcal{C}:u^4)\subseteq(\mathcal{C}:u^5)
\end{equation}
For $i=1,\cdots,5$ the projections of $(\mathcal{C}:u^i)$ over the field $K$ are denoted by $Tor_i(\mathcal{C})=(\overline{\mathcal{C}:u^i})$ and are called the {\it torsion codes} associated to the code $\mathcal{C}$.
\newline
The following theorem presents some bounds on the edit distance for the cyclic codes defined above.
\begin{thm}
Let $\mathcal{C}=\langle f_0,uf_1,u^2f_2,u^3f_3,u^4f_4,u^5f_5\rangle$ be a cyclic code over $R$ of odd length. Then the minimum edit distance $d_c$ of $\mathcal{C}$ satisfies the following inequalities.
\newline
$(i)$ $d_c(\mathcal{C})=min\{d_c(Tor_i(\mathcal{C}))\}\leq min\{d_H(Tor_i(\mathcal{C}))\};$
\newline
$(ii)$ $d_c(\mathcal{C})\leq min\{deg(f_i)\}+1;$
\newline
$(iii)$ $d_c(\mathcal{C}) \leq n-rank(\mathcal{C})+1.$
\end{thm}
\pf
From \cite[Lemma 5.1]{MDS} and Proposition ~\ref{prop:edit}, we have that $d_c(\mathcal{C})=min\{d_c(Tor_i(\mathcal{C}))\}\leq min\{d_H(Tor_i(\mathcal{C}))\}.$ Assertion $(ii)$ comes from the fact that the code $Tor_i(\mathcal{C})$ and $Tor_0(\mathcal{C})$ are binary cyclic codes, and satisfy $\langle f_i \rangle \subset \mathcal{C} .$ The dimension of $Tor_i(\mathcal{C})$ is $n-deg(f_i).$  By the well-known Singleton bound, we have $d_c(\mathcal{C})\leq min\{deg(f_i)\}+1.$ Assertion $(iii)$ follows from Proposition ~\ref{prop:edit} using the Singleton bound.
\qed
\subsection{DNA Cyclic Codes }
Now,  we firstly introduce a DNA cyclic code, more precisely,  the construction of a $[3n,d]$-DNA cyclic code.
Set $\mathbb{I}(x):=(x^n-1)/(x-1)$ and  $\alpha(u):=u^5+u^4+u^3+u^4+u+1$.
\begin{defi}
\label{defi:DNA}
Let  $1\leq D\leq 3n-1$ be a positive real number. Then a cyclic code $\mathcal{C}$  of the length $n$ over $R$ is called an  $[n,D]$-cyclic DNA code  if the following conditions hold:
\begin{itemize}
\item[(i)] $\mathcal{C}$ is cyclic code, i.e., $\mathcal{C}$ is an ideal in  $R_{n}=R[x]/(x^{n}-1);$
\item[(ii)] for any codeword $x\in C$, we have $(x)^{rc}\neq (x)$ and $(x)^{rc}\in\mathcal{C};$
\item[(iii)]$d_c(x,y) \leq D $ for any $x,y\in C$.
\end{itemize}
\end{defi}
Condition (ii) given in Definition~\ref{defi:DNA} shows that the defined DNA cyclic codes are reverse-complement cyclic codes.
\begin{defi}
Let $f(x)\in R[x]$, denote $f(x)^{*}=x^{deg(f)}f(\frac{1}{x})$ the reciprocal polynomial of $f(x)$. The polynomial $f$ is said to be self-reciprocal if $f(x)=f^*(x).$
\end{defi}
\begin{lem}~\label{lem:recip}(\cite{GG12})
Let  $f(x)$ and  $g(x)$  be a polynomials in $R[x]$ with $deg(f(x))\geq deg(g(x))$. Then the following conditions hold:
\begin{itemize}
\item[(i)]$[f(x)g(x)]^{*}=f(x)^{*}g(x)^{*};$
\item[(ii)]$[f(x)+g(x)]^{*}=f(x)^{*}+x^{deg (f)-deg (g)}g(x)^{*}.$
\end{itemize}
\end{lem}
\begin{thm}
 Let $\mathcal{C}$ be a cyclic code of odd length $n$ over $R$, and that  $\mathcal{C}$ is reverse-complement. Then we have:
\begin{itemize}
\item[(i)] $\mathcal{C}$ contains all the codewords of the form $\alpha(u)\mathbb{I}(x)$.
\item[(ii)] $\mathcal{C}=\langle f_0,uf_1,u^2f_2,u^3f_3,u^4f_4,u^5f_5\rangle$; such that all $f_i$  self-reciprocal.
\end{itemize}
\end{thm}
\pf
\begin{itemize}
\item[(i)] Assume $\mathcal{C}=\langle f_0,uf_1,u^2f_2,u^3f_3,u^4f_4,u^5f_5\rangle$ is a cyclic code of odd length $n$ over $R.$ Since the zero codeword is an element in $\mathcal{C}$ then its WCC is also a codeword of $\mathcal{C},$ i.e.;

     $$(\hat{0},\hat{0},\cdots,\hat{0})=(\alpha(u),\alpha(u),\cdots,\alpha(u))$$
$$=\alpha(u)(1,1,\cdots,1)=\alpha(u)(x^n-1)/(x-1)\in \mathcal{C}.$$

\item[(ii)]
In the second part we show that $f_i^*(x)=f_i(x)$ for $ 0\le i \le 5 $.
\newline
Let $f_0(x)=a_0+a_1x+\cdots+a_{m-1}x^{m-1}+a_mx^m$ where $f_0/(x^n-1)$ in $\mathbb{F}_2[x].$ One can assume that $a_0=a_m=1.$ So that $f_0(x)=1+a_1+\cdots+a_{m-1}x^{m-1}+x^m$. Suppose that $f_0(x)$ corresponds to the vector  $(1,a_1,\cdots,1,0,0\cdots,0)$ and the  reverse-complement of  $0$ in $R$ is $\alpha(u)$ then $f_{0}^{rc}(x)=\alpha(u)(1+x+\cdots+x^{n-m-2})+(\alpha(u)+1)x^{n-m-1}+\hat{a}_{m-1}x^{n-m}+\cdots+
\hat{a}_1x^{n-2}+(\alpha(u)+1)x^{n-1})\in\mathcal{C}$.
Now, since $\mathcal{C}$ is a linear code, we get

$$f_0(x)^{rc}+\alpha(u)(\frac{x^n-1}{x-1})\in \mathcal{C}.$$
This implies that

$$x^{n-m-1}+(\hat{a}_{m-1}+\alpha(u))x^{n-m}+\cdots+(\hat{a}_1+\alpha(u))x^{n-2}+x^{n-1}$$
$$=x^{n-m-1}[1+(\hat{a}_{r-1}+\alpha(u))x+\cdots+(\hat{a}_1+\alpha(u))x^{m-1}+x^m]\in \mathcal{C}.$$

  Multiplying the last polynomial by  $x^{m+1}$ in  $R[x]/(x^n-1)$, we obtain:

$$1+(\hat{a}_{m-1}+\alpha(u))+\cdots+(\hat{a}_1+\alpha(u))x^{m-1}+x^m\in \mathcal{C}.$$

By  Equation (~\ref{eq:hat}) we see that  $\hat{a}+\alpha(u)=a$. Therefore,  we obtain:

  $$f_0^*(x)=1+a_{m-1}x+\cdots+a_1x^{m-1}+x^m\in C.$$

   Consequently, we have

$$f_0^*(x)=f_0k_0+uf_1k_1+\cdots+u^5f_5k_5,$$
 where $f_i$ and  $k_i$ are all in  $\mathbb{F}_2[x]$. Multiplying both sides of this equality by $u^5$ gives
 $$u^5f_0^*(x)=u^5k_0(x)f_0(x).$$
Now,  since $f_0^*(x),f_0(x)\in\mathbb{F}_2[x]$ have the same degree, leading coefficient and  constant term, one necessary have $k_0(x)=1.$ Consequently, $f_0(x)$ is self-reciprocal. The same argument can be used for $f_1,f_2,f_3,f_4$ and $f_5$ as well.
\end{itemize}
\qed

In the following, we are interested in providing sufficient conditions for a given code $\mathcal{C}$ to be  reverse-complement.

\begin{thm}
\label{thm:reverse2}
Assume that $\mathcal{C}=\langle f_0,uf_1,u^2f_2,u^3f_3,u^4f_4,u^5f_5\rangle$ is a cyclic code of odd length  $n$  over $R$ with $f_5|f_4|f_3|f_2|f_1|f_0|x^n-1\in\mathbb{F}_2[x].$ If $\alpha(u)\mathbb{I}(x)\in C$ and $f_i(x)$ are self-reciprocal, then $\mathcal{C}$ is a reverse-complement code.
\end{thm}
\pf
Let $c(x)$ be a codewords  in $\mathcal{C}$, we have to prove that $c(x)^{rc}\in\mathcal{C}$. Since $\mathcal{C}=\langle f_0,uf_1,u^2f_2,u^3f_3,u^4f_4,u^5f_5\rangle$ there exist
 $\alpha_i(x)\in R[x]$ ($i\in\{1,\cdots,5\}$) such that
 $$c(x)=f_0\alpha_0+uf_1\alpha_1+u^2f_2\alpha_2+u^3f_3\alpha_3+u^4f_4\alpha_4+u^5f_5\alpha_5.$$
Applying the reciprocal and using first Lemma \ref{lem:recip}, and next the fact that $f_0(x),f_1(x),f_2(x),f_3(x),f_4(x)$ and
$f_5(x)$ are self-reciprocal, we obtain $c^*(x)= (f_0\alpha_0)^*+(uf_1\alpha_1)^*x^{m_1}+(u^2f_2\alpha_2)^*x^{m_2}
+(u^3f_3\alpha_3)^*x^{m_3}+(u^4f_4\alpha_4)^*x^{m_4}+(u^5f_5\alpha_5)^*x^{m_5}$, proving that $c(x)^*$ is in $\mathcal{C}$. Since $\mathcal{C}$ is cyclic, $x^{n-t-1}c(x) = c_0x^{n-t-1} + c_1x{n-t} +\cdots+c_tx^{n-1}\in\mathcal{C}.$ It was also assumed that $\alpha(u)+ \alpha(u)x+\cdots+\alpha(u)x^{n-1}\in\mathcal{C}$, which leads to
$$\alpha(u)+\alpha(u)x+\cdots+\alpha(u)x^{n-1}+c_0x^{n-t-1} + c_1x{n-t} +\cdots+c_tx^{n-1}\in\mathcal{C}. $$
This is equal to
$\alpha(u)+\alpha(u)x+\cdots+\alpha(u)x^{n-t-2}+(\alpha(u)+c_0)x^{n-t-1}+\cdots+(\alpha(u)+c_t)x^{n-1}=\alpha(u)+\alpha(u)x\cdots+
\alpha(u)x^{n-t-1}+\cdots+\hat{c}_0x^{n-t-1}+\cdots+\hat{c}_tx^{n-1},$ which is precisely $(c^*(x)^{rc})^*=c(x)^{rc}\in \mathcal{C}.$
\qed
\begin{thm}
Assume that $\mathcal{C}=\langle f_0,uf_1,u^2f_2,u^3f_3,u^4f_4,u^5f_5\rangle$ be a cyclic code of even length  $n=m2^s$  over $R$ such that $f_i|f_0$ in $\mathbb{F}_2[x].$ If $\alpha(u)\mathbb{I}(x)\in C$ and $f_i(x)$ ($i\in\{1,\cdots,5\}$) are self-reciprocal. Then $\mathcal{C}$ is a reverse-complement code.
\end{thm}
\pf
The proof can be done in a similar  manner than the one of Theorem~\ref{thm:reverse2}.
\qed
\begin{cor}
Let $C$ be a cyclic code of length $n=m2^s$, $s\geq 0.$ If $\alpha (u)\mathbb{I}(x)\in \mathcal{C}$ and there exists an integer $i$ such that
 \begin{equation}
 \label{eq:fac2}
2^i\equiv -1[m].
 \end{equation}
 Then $\mathcal{C}$ is a reverse-complement code.
\end{cor}
\pf
The proof is similar to the proof of Corollary 4.14 in \cite{GGS}.
\qed
\newline
\begin{defi}
For a cyclic code $\mathcal{C}=\langle f_0,uf_1,u^2f_2,u^3f_3,u^4f_4,u^5f_5\rangle$, we define the sub-code  $\mathcal{C}_{u^2}$ consisting of all codewords in $\mathcal{C}$ that are a multiple of $u^2$.
\end{defi}
\begin{lem}
\label{lem:triple}
With the previous notations we have:
\begin{itemize}
\item [(i)]
$\phi(u^2R)=\{GGG,AGT,CGT,TGT,GAT,CAG,TAC,ATC,GTC,TCC,CTC,ACC,\\CTC,TCC,GGT,CAT\},$
\newline
$\phi(u^3R)=\{GGG,TGT,CAG,TAC,CTC,GCC,AGT,TTA\}$ and
\newline
$\phi(u^4R)=\{GGG,TAC,CTC,TGT\}.$
\item [(ii)] If $\mathcal{C}=\langle f_0,uf_1,u^2f_2,u^3f_3,u^4f_4,u^5f_5\rangle$ is the cyclic code of length  $n$ over $R$. Then $\mathcal{C}_{u^2}=\langle u^2f_5\rangle $ over the alphabet $\phi(u^2R).$
\end{itemize}
\end{lem}
\pf
  The part (i) is obtained by a simple calculation.
 \newline
 For the part (ii), assume that $\mathcal{C}=\langle f_0,uf_1,u^2f_2,u^3f_3,u^4f_4,u^5f_5\rangle$. Since $f_5|f_4|f_3|f_2|f_1|f_0|x^n-1$, then we obtain $\langle u^2f_5\rangle\subset \mathcal{C}_{u^2}.$ Conversely, assume that  $c(x)\in C$ such that  $c(x)=\alpha_0(x)f_0(x)+u\alpha_1(x)f_1(x)+u^2\alpha_2(x)f_2(x)+u^3\alpha_3(x)f_3(x)+u^4f_4+u^5f_5$ for all $\alpha_i\in\mathbb{F}_2[x]$. If  $c(x)$ is a multiple of $u^2$ then  $x^n-1$ divides $\alpha_0(x)f_0(x)$ and $x^n-1$ divides $\alpha_1(x)f_1(x)$. Hence, $c(x)=u^2\alpha_2(x)f_2(x)+u^3\alpha_3(x)f_3(x)+u^4\alpha_4(x)f_4(x)+u^5\alpha_5(x)f_5(x)$. Therefore, $\mathcal{C}_{u^2}\subset\langle u^2f_5\rangle$. Consequently $\mathcal{C}_{u^2}=\langle u^2f_5\rangle$, which completes the proof.
\qed
\begin{rem}
The cyclic DNA codes which are obtained in the Lemma \ref{lem:triple} are stable across the error in the DNA strands by the usage of the codons, see \cite{Marc}.
\newline
Any codeword of sub-code $\mathcal{C}_{u^2}=\langle u^2f_5\rangle$ over $\phi(u^2R)$ contains the nucleotide $C$ and $G$.
This is an interesting thermodynamic property of the DNA strand. For its importance please see \cite{GGS}.
\end{rem}
 \begin{ex}
We have that  $x^7-1=(x-1)(x^3+x+1)(x^3+x^2+1)=f_0f_1f_2$ in $\mathbb{F}_2[x].$ The Table 2 represents the cyclic DNA code with minimal Hamming distance.
 \begin{table}[h]
\caption{ Cyclic DNA Codes with Minimal Hamming Distance}
\begin{center}
\begin{tabular}{|c|c|c|}
  \hline
  Number of the Code & Generators & Type of the Code \\
  \hline
  1 & $<u^2f_0>$ & $(7,4096,2)$ \\
  2 & $<u^2f_1>$ & $(7,256,3)$  \\
  3 & $<u^2f_2>$ & $(7,256,3)$\\
  4 & $<u^2f_1f_2>$ & $(7,4,7)$ \\
  5 & $<u^2f_0f_1>$ & $(7,64,4)$  \\
  6 & $<u^2f_0f_2>$ & $(7,64,4)$ \\
  \hline
\end{tabular}
 \end{center}
\end{table}
\begin{table}[h]
\caption{A DNA Code with Minimal Edit Distance Obtained from  $\mathcal{C}=\langle u^4f_0f_1\rangle$ of the Type $(7,64,6)$}
\begin{center}
\tiny{
\begin{tabular}{|c|c|}
  \hline
  GGGGGGGGGGGGGGGGGGGGG&CCCCCCCCCCCCCCCCCCCCC \\
  CTCGGGCTCCTCCTCGGGGGG&GAGCCCGAGGAGGAGCCCCCC \\
  GGGCTCGGGCTCTGTTGTTGT&CCCGAGCCCGAGACAATAACA \\
  TGTGGGCTCGGGCTCTGTTGT&ACACCCGAGCCCGAGACAACA \\
  TGTTGTGGGCTCGGGCTCTGT&ACAACACCCGAGCCCGAGACA \\
  TGTTGTTGTGGGCTCGGGCTC&ACAACAACACCCGAGCCCGAG \\
  CTCTGTTGTTGTGGGCTCGGG&GAGACAACAACACCCGAGCCC \\
  GGGCTCTGTTGTTGTGGGCTC&CCCGAGACAACAACACCCGAG \\
  TATGGGTATTATTATGGGGGG&ATACCCATAATAATACCCCCC \\
  GGGTATGGGTATTATTATGGG&CCCATACCCATAATAATACCC \\
  GGGGGGTATGGGTATTATTAT&CCCCCCATACCCATAATAATA \\
  TATGGGGGGTATGGGTATTAT&ATACCCCCCATACCCATAATA \\
  TATTATGGGGGGTATGGGTAT&ATAATACCCCCCATACCCATA \\
  TATTATTATGGGGGGTATGGG&ATAATAATACCCCCCATACCC \\
  GGGTATTATTATGGGGGGTAT&CCCATAATAATACCCCCCATA \\
  TGTGGGTGTTGTTGTGGGGGG&ACACCCACAACAACACCCCCC \\
  GGGTGTGGGTGTTGTTGTGGG&CCCACACCCACAACAACACCC \\
  GGGGGGTGTGGGTGTTGTTGT&CCCCCCACACCCACAACAACA \\
  TGTGGGGGGTGTGGGTGTTGT&ACACCCCCCACACCCACAACA \\
  TGTTGTGGGGGGTGTGGGTGT&ACAACACCCCCCACACCCACA \\
  TGTTGTTGTGGGGGGTGTGGG&ACAACAACACCCCCCACACCC \\
  GGGTGTTGTTGTGGGGGGTGT&CCCACAACAACACCCCCCACA \\
  CTCGGGCTCTGTTGTTGTGGG&GAGCCCGAGACAACAACACCC \\
  GGGCTCGGGCTCTGTTGTTGT&CCCGAGCCCGAGACAACAACA \\
  TGTGGGCTCGGGCTCTGTTGT&ACACCCGAGCCCGAGACAACA \\
  TGTTGTGGGCTCGGGCTCTGT&ACAACACCCGAGCCCGAGACA \\
  TGTTGTTGTGGGCTCGGGCTC&ACAACAACACCCGAGCCCGAG \\
  CTCTGTTGTTGTGGGCTCGGG&GAGACAACAACACCCGAGCCC \\
  GGGCTCTGTTGTTGTGGGCTC&CCCGAGACAACAACACCCGAG \\
  GGGGGGCTCGGGCTCCTCCTC&CCCCCCGAGCCCGAGGAGGAG \\
  CTCGGGGGGCTCGGGCTCCTC&GAGCCCCCCGAGCCCGAGGAG \\
  CTCCTCGGGGGGCTCGGGCTC&GAGGAGCCCCCCGAGCCCGAG \\
  \hline
\end{tabular}
}
\end{center}
\end{table}
\end{ex}
\subsection{Binary Image of  DNA Codes}
In this Section we will define a Gray map which allows us to translate the properties of the suitable DNA codes for DNA computing to the binary cases.
The definition of the Gray map $\varphi$ from $R$ to $\mathbb{F}_2,$ for each element of $\mathbb{F}_2$ is expressed as
$$\varphi(a_0+a_1u^1+a_2u^2+a_3u^3+a_4u^4+a_5u^5)=(a_0,a_1,a_2,a_3,a_4,a_5),$$
where $a_i\in\mathbb{F}_2.$
We define Lee weight over the ring $R$ by
$$w_{Lee}(a_0+a_1u^1+a_2u^2+a_3u^3+a_4u^4+a_5u^5)=\sum_{i=0}^{i=5}a_i.$$
 The Lee distance $d_L(x,y)$ between the vector $x$ and $y$ is $w_{Lee}(x-y).$ According to the definition of the Gray map, it is  easy to verify that the image of a linear code over $R$ by $\varphi$ is a binary linear code. We can obtain the binary image of the DNA code by the map $\varphi$ and the map $\phi.$ In Table 4 we give the binary image of the codons. The binary image of DNA code  resolved the problem of the construction of DNA codes with some properties, see \cite{ImageB}.
\begin{table}[h]
\caption{ Binary Image of the Codons}
\begin{center}

\begin{tabular}{cccccccc}
  \hline
  GGG & 000000 & CCC & 111111 & TAT & 000001 & ATA & 111110\\
  GGA & 011111 & CCT & 100000 & TAC & 100001 & ATG & 011110\\
  GGC & 101111 & CCG & 010000 & TAA & 010001 & ATT & 101110\\
  GGT & 001111 & CCA & 110000 & TAG & 110001 & ATC & 001110\\
  AGG & 110111 & TCC & 001000 & CAT & 001001 & GTA & 110110\\
  AGA & 010111 & TCT & 101000 & CAC & 011001 & GTG & 100110\\
  AGC & 100111 & TCG & 011000 & CAA & 011001 & GTT & 100110\\
  AGT & 000111 & TCA & 111000 & CAG & 111001 & GTC & 000110\\
  CGG & 111011 & GCC & 000100 & AAT & 000101 & TTA & 111010\\
  CGA & 011011 & GCT & 100100 & AAC & 100101 & TTG & 011010\\
  CGC & 101011 & GCG & 010100 & AAA & 010101 & TTT & 101010\\
  CGT & 001011 & GCA & 110100 & AGG & 110101 & TCC & 001010\\
  TGG & 110011 & ACC & 001100 & GAT & 001101 & CTA & 110010\\
  TGA & 010011 & ACT & 101100 & GAC & 101101 & CTG & 010010\\
  TGC & 100011 & ACG & 011100 & GAA & 011101 & CTT & 100010\\
  TGT & 000011 & ACA & 111100 & GAG & 111101 & CTC & 000010\\
  \hline
\end{tabular}
\end{center}
\end{table}

The following property of the binary image of the  DNA codes comes from the definition.
\begin{lem}
The Gray map $\varphi$ is a linear distance preserving
\begin{center}
($R^n$, Lee distance)$\rightarrow$(${\mathbb{F}_2}^{6n}$, Hamming distance).
\end{center}
Further if $\mathcal{C}$ is a cyclic DNA code over $R$ then $\varphi(\mathcal{C})$ is a binary quasi-cyclic DNA code of the length $6n$ and of  index $6$.
\end{lem} \label{lem:image}
\pf
Let $\mathcal{C}$ be a cyclic DNA code of length $n$ over $R.$ Hence $\varphi(\mathcal{C})$ is a set of length $6n$ over the alphabet $\mathbb{F}_2$ which is a quasi-cyclic code of  index 6. It is easy to verify that the Gray map is a linear distance preserving.
\qed
\section{Skew Cyclic DNA Codes over $\tilde {R}=\mathbb{F}_2+v\mathbb{F}_2=\{0,1,v,v+1\}$}

\subsection{Notation and Prelimenaries}
The ring considered in this section is the non-commutative ring $\tilde {R}[x;\theta]$. The structure of this non-commutative ring depends on the element of the commutative ring $\tilde {R}=\{0,1,v,v+1\}$, where $v^2=v$ and the automorphism $\theta$ on $\tilde {R}$, defined by $\theta(0)=0,$ $\theta(1)=1,$ $\theta(v)=v+1,$ $\theta(v+1)=v$. Note that $\theta^2(a)=\theta(\theta(a))=a$ for all $a\in \tilde {R}$. This implies that $\theta$ is a ring automorphism of order 2.
The skew polynomial ring $\tilde {R}[x;\theta]$ is the set of polynomials $\tilde {R}[x;\theta]=\{a_0+a_1x+a_2x^2+\cdots+a_nx^n|a_i\}$, where addition is the usual polynomial addition and the multiplication is not commutative over $\tilde {R}$, denote by  $*$ and  defined by the basic rule $(ax^i)*(bx^j)=a\theta^i(b)x^{i+j}$
and the distributive and the associative laws.
\newline
We have one-to-one map $\psi$ between the elements of $\tilde {R}$ and the DNA nucleotide  base $\{A,T,C,G\}$ given by
$0\rightarrow G$, $v \rightarrow C$, $v+1 \rightarrow T$ and $1\rightarrow A$. A simple
verification give that for all $x\in \tilde {R},$ we have
 \begin{equation}
 \label{eq:hat}
 \theta(x)+\theta(\hat{x})=v+1.
 \end{equation}

In the following, we only consider codes with even length.
\begin{defi}
Let $\tilde {R}=\mathbb{F}_2+v\mathbb{F}_2=\{0,1,v,v+1\}$ be a ring where $v^2=v$ and the automorphism $\theta$ defined previously. A subset $\tilde{\mathcal C}$ of $\tilde {R}^n$ is called a skew cyclic code ($\theta$-cyclic code) of the length $n$ if
\begin{enumerate}
  \item  $\tilde{\mathcal C}$ is a $R$-submodule  of $R^n$ and
  \item if $c=(c_0,c_1,\cdots,c_{n-1})\in \tilde{\mathcal C}$ then $(\theta(c_{n-1}),\theta(c_0),\cdots,\theta(c_{n-2}))\in\tilde{\mathcal C}$
\end{enumerate}
\end{defi}
 The ring $\tilde {R}_n=\tilde {R}[x;\theta]/(x^n-1)$ denotes the quotient ring of $\tilde {R}[x;\theta]$ by the (left) ideal $(x^n-1)$. Let $f(x)\in \tilde {R}_n$ and $r(x)\in \tilde {R}[x;\theta],$  we define  the  multiplication from left as:
\begin{equation}
 \label{eq:left}
r(x)*(f(x)+(x^n-1))=r(x)*f(x)+(x^n-1)
 \end{equation}
for any  $r(x)\in \tilde {R}[x;\theta].$
Define  a map as follows
$$\xi:  \tilde {R}^n\rightarrow \tilde {R}[x;\theta]/(x^n-1)$$
$$(c_0,c_1,\cdots,c_{n-1})\rightarrow c_0+c_1x+c_2x^2+\cdots+c_{n-1}x^{n-1}.$$
It is clear that $\xi$ is an $\tilde {R}$-module isomorphism map. This implies that each element $(c_0+c_1\cdots+c_{n-1})\in \tilde {R}^n$ can be identified by the polynomial $c(x)=c_0+c_1x+c_2x^2+\cdots+c_{n-1}x^{n-1}\in \tilde {R}_n.$
\begin{lem} (\cite[Lemma 1]{A2})
If $n$ is even, and $x^n-1=g(x)*f(x)$ in $\tilde {R}[x;\theta],$ then $x^n-1=g(x)*f(x)=f(x)*g(x)$
\end{lem}
The following proposition gives any structures of the skew cyclic codes over $\tilde R_{n}$.
\begin{prop}(\cite[ Corollary 3]{A2})
\label{cor:struct1}
Let $\tilde{\mathcal C}$ be a skew cyclic code in $\tilde R_{n}$. Then
\begin{enumerate}
  \item  If a polynomial $g(x)$ of least degree in $\tilde{\mathcal C}$  is a monic  then  $\tilde{\mathcal C}=(g(x)),$  where $g(x)$ is (skew) right divisor of $x^n-1$.
  \item  If $\tilde{\mathcal C}$  contains some monic polynomial but no polynomials $f(x)$ least degree in $\tilde{\mathcal C}$  is  monic. Then $\tilde{\mathcal C}=(f(x),g(x)),$ where $g(x)$ is a monic polynomial of least degree in $\tilde{\mathcal C}$ and $f(x)=vf_1(x)$ or $f(x)=(v+1)f_1(x)$ for some polynomial binary $f_1(x).$
  \item  If $\tilde{\mathcal C}$  does not contain any monic polynomials. Then $\tilde{\mathcal C}=(f(x))$ where $f(x)=vf_1(x)$ or $f(x)=(v+1)f_1(x)$ and $f_1(x)$ is a binary polynomial that divides $x^n-1$.
\end{enumerate}

\end{prop}
Now, we are interesting in constructions of $[n,d]$-skew cyclic DNA  codes. To this end, we start by defining such  codes.

\begin{defi}
Let  $1\leq d\leq n-1$ be a positive real number. A skew cyclic code $\tilde{\mathcal C}$ over $\tilde R$ is  said to be a  $[n,d]$-skew cyclic DNA code  if the following conditions hold.
\begin{enumerate}
\item $\tilde{\mathcal C}$ is a skew cyclic code, that is, $\tilde{\mathcal C}$ is a $\tilde R$-submodule  of  $\tilde R_{n}$
\item  for any codeword $X\in \tilde{\mathcal C}$: $(X)^{rc}\neq (X)$ and $(X)^{rc}\in
(C)$
\item $d_H(X,Y)\leq d$ for any $X,Y\in C$.
\end{enumerate}
\end{defi}
\subsection{The Reverse-Complement Skew Cyclic Codes over  $\tilde {R}$}
In this subsection, we give conditions on  the existence of the reverse-complement cyclic
codes of the even length $n$ over the ring $\tilde {R}.$

Let $v=(a_0,a_1,\cdots,a_{n-2},a_{n-1})$ be a vector in $\tilde {R}_n$, the reverse of the vector $v$ is $v^r=(a_{n-1},a_{n-2},\cdots,a_1,a_0)$.
Let $f(x)$ is the polynomial correspond of the vector$v$ such that $f(x)=a_0+a_1x+\cdots+a_{n-1}x^{n-1}$, we find the polynomial correspond of the vector $v^r$ in $R[x;\theta]$, we multiply  the right of the polynomial $f(x^{-1})$ by $x^{n-1}$, we obtain: $f(x^{-1})x^{n-1}=a_0x^{n-1}+a_1\theta(1)x^{n-2}+\cdots+a_{n-2}\theta^{n-2}(1)x+a_n\theta^{n-1}(1)=a_{n-1}+a_{n-2}x+\cdots+a_1x^{n-2}+a_0x^{n-1}.$
This polynomial corresponds of the vector $v^r$ denoted by $f^r(x)$ (see Section 1).
\begin{defi}
Let  $f(x)^{*}=f(x^{-1})*x^{deg(f)}$ be the reciprocal polynomial of a given $f(x)$ in $\tilde {R}[x;\theta]$.
The polynomial $f$ is called self-reciprocal if $f$ coincides with $f^*$.
\end{defi}\label{def:rev1}
\begin{ex}
Let $f$ be a polynomial in $\tilde {R}[x;\theta]$ such that,
\newline
$f(x)=x^3+vx^2+(v+1)x+v$, this polynomial represented the DNA sequence $X(ACTC)$, we find  the reverse of the sequence $X$ by $f^*(x).$
\newline
$f^*(x)=f(x^{-1})x^3=1\theta^3(1)+v\theta^2(1)x+(v+1)\theta(1)x^2+v\theta^0(1)x^3=vx^3+(v+1)x^2+vx+1.$  Then we have the reverse of the DNA sequence of $X$ is $(CTCA).$
\end{ex}
Notice that the definition of reciprocal polynomial over $\tilde {R}[x,\theta]$ is being different from the one defined over a commutative ring. Indeed,  in the non-commutative ring $\tilde {R}[x,\theta]$, we use the right multiplication over the automorphism $\theta$ and the multiplication over $\tilde {R}[x,\theta]$.
\begin{lem}~\label{lem:recip}
Let  $f(x)$ and  $g(x)$  be polynomials in $\tilde {R}[x,\theta]$ with $deg(f(x))\geq deg(g(x))$. Then the following assertions hold:
\begin{itemize}
\item[(i)]$[f(x)g(x)]^{*}=f(x)^{*}g(x)^{*}$
\item[(ii)]$[f(x)+g(x)]^{*}=f(x)^{*}+g(x)x^{deg (f)-deg (g)}$
\end{itemize}
\end{lem}
\pf
Assertion $(i)$ assume that $f(x)= \sum_{i=0}^na_ix^i$ and $g(x)= \sum_{j=0}^pb_jx^j$, such that the $deg(f)\geq deg(g),$ we have $f(x)g(x)=\sum_{k=0}^{n+p}\sum_{i=0}^ka_i\theta^i(b_{k-i})x^k$, by Definition \ref{def:rev1}, we have that $(f(x)g(x))^*=(\sum_{k=0}^{n+p}\sum_{i=0}^ka_i\theta^i(b_{k-i})x^{-k})x^{n+p}$. Then we have $(f(x)g(x))^*=\sum_{k=0}^{n+p}\sum_{i=0}^ka_i\theta^i(b_{k-i})x^{n+p-k}$.
 By the Definition \ref{def:rev1} $f(x)^*=\sum_{i=0}^na_i\theta^i(1)x^{n-i}=\sum_{i=0}^na_ix^{n-i}$ and $g(x)^*=\sum_{j=0}^pb_j\theta^j(1)x^{p-j}=\sum_{j=0}^pb_jx^{p-j}$. Consequently we have $f(x)^*g(x)^*=\sum_{k=0}^{n+p}\sum_{i=0}^ka_i\theta^i(b_{k-i})x^{n+p-k}.$
 Hence the result follows.
\newline
Assertion $(ii)$, by Definition \ref{def:rev1}, we have
\newline
$(f(x)+g(x))^*=(f+g)^*(x)=((f+g)(x^{-1}))x^{deg(f)}\\=(f(x^{-1})+g(x^{-1}))x^{deg(f)}=(f(x^{-1})x^{deg(f)}+g(x^{-1}x^{deg(f)})\\=(f^*(x)+g(x^{-1}x^{deg(f)})
=f^*(x)+g(x^{-1})x^{deg(g)}x^{deg(f)-deg(g)}\\=f^*(x)+g^*(x)x^{deg(f)-deg(g)}.$ The result follows.\\
\qed

In the following we are interested in providing necessary conditions for $\tilde{\mathcal C}$ to be a reverse-complement code.
\begin{thm}
\label{thm:reves1}
Let $\tilde{\mathcal C}=(f(x))$ be a skew cyclic code  in $\tilde {R}_n$, where $f(x)$ is monic polynomial of minimal degree.  If $\tilde{\mathcal C}$ is reverse-complement then  the polynomial $f(x)$ is self-reciprocal and $v(x^n-1)/(x-1)\in \tilde{\mathcal C}$.
\end{thm}

\pf
Let $\tilde{\mathcal C}=(f(x))$ be a skew cyclic code over $\tilde {R}$, where $f(x)$ is monic polynomial of minimal degree in $\tilde{\mathcal C}.$ We know that $(0,0,\cdot\cdot\cdot,0)\in\tilde{\mathcal C},$ since $\tilde{\mathcal C}$ is reverse-complement then$(0,0,\cdot\cdot\cdot,0)^{rc}\in\tilde{\mathcal C}$ i,e.;
$(\hat{0},\hat{0},\cdot\cdot\cdot,\hat{0})=(v,v,\cdot\cdot\cdot,v)\in\tilde{\mathcal C}$, this vector correspond of the polynomial
$v+vx+\cdot\cdot\cdot+vx^{n-1}=v(x^n-1)/(x-1)\in\tilde{\mathcal C}.$ We have that $f(x)$ is monic polynomial of minimal degree  in $\tilde{\mathcal C},$ where $f(x)=1+a_1x+\cdot\cdot\cdot+x^t$, the vector correspond to the polynomial $f(x)$ is $(1,a_1,\cdot\cdot\cdot,0,0,\cdot\cdot\cdot,0),$ since $\tilde{\mathcal C}$ is reverse-complement and linear, then $(1,a_1,\cdot\cdot\cdot,0,0,\cdot\cdot\cdot,0)^{rc}\in\tilde{\mathcal C}$, i.e.,
\newline
$f^{rc}(x)=v+vx+\cdot\cdot\cdot+vx^{n-t-2}+(v+1)x^{n-t-1}+a_{t-1}x^{n-t}+\cdot\cdot\cdot+a_1x^{n-2}+vx^{n-1}
\\=f^{rc}(x)+v(x^n-1)/(x-1)\in\tilde{\mathcal C}.$ This implies that
\newline
$x^{n-t-1}+(\hat{a_{t-1}}+v)x^{n-t}+\cdot\cdot\cdot+(\hat{a_1}+v)x^{n-2}+x^{n-1}\in\tilde{\mathcal C}.$
\newline
Multiplying on the right by $x^{t+1-n},$ we obtain,
\newline
$(1+(\hat{a}_{t-1}+v)\theta(1)x+\cdot\cdot\cdot+(\hat{a}_1+v)\theta^{t-1}(1)x^{t-1}+\theta^t(1)x^{t})x^{t-n-1}\in\tilde{\mathcal C}.$
\newline
Hence,
$(1+(\hat{a}_{t-1}+v)x+\cdot\cdot\cdot+(\hat{a}_1+v)x^{t-1}+x^{t})\in \tilde{\mathcal C},$ which implies
(thanks to  Equation (~\ref{eq:hat}))  that
$f^*(x)=1+a_{t-1}x+\cdot\cdot\cdot\cdot+x^t\in\tilde{\mathcal C}.$
 Since $\tilde{\mathcal C}=(f(x))$, there exists $q(x)\in R[x, \theta]$ such that $f^*(x)=q(x)f(x)$, one necessary have $q(x)=1,$ that is $f^*(x)=f(x).$
\newline

\qed

\begin{thm}
\label{thm:reves2}
Let $\tilde{\mathcal C}=(vf_1(x))$ be a skew cyclic code in $\tilde {R}_n$, where $f_1(x)$ is a monic binary polynomial of lowest degree with $f_1(x)|(x^n-1)$. If $\tilde{\mathcal C}$ is a reverse-complement code then  $f_1(x)$ is self-reciprocal.
\end{thm}
\pf
Suppose $f_1(x)=1+a_1x+a_2x+\cdot\cdot\cdot+x^r$ is a binary polynomial, the vector correspond of $f(x)$ is $v=(1,a_1,\cdot\cdot\cdot,a_{r-1},1,0,0,0,\cdot\cdot\cdot,0,0).$ Hence $v^{rc}=(\hat{0},\hat{0},\hat{0},\cdot\cdot\cdot,\hat{0},\hat{1},\hat{a}_{r-1},\cdot\cdot\cdot,\hat{a}_1,\hat{1}).$ This vectors correspond of the polynomial
\newline
$f_1^{rc}(x)=v+vx+\cdot\cdot\cdot+vx^{n-r-2}+(v+1)x^{n-r-1}+\hat{a}_{n-r}x^{n-r}+\cdot\cdot\cdot+\hat{a}_1x^{n-2}+(v+1)x^{n-1}
=f_1^{rc}+v(x^n-1)(x-1)$. Since $\tilde{\mathcal C}$ is a linear code then, $f_1^{rc}+v(x^n-1)(x-1)\in\tilde{\mathcal C}.$ Therefore
\newline
$x^{n-r-1} +(\hat{a}_{r-1}+v)x^{n-1}+\cdot\cdot\cdot+(\hat{a}_1+v)x^{n-2}+x^{n-1}\in\tilde{\mathcal C},$  we multiply  by  $x^{-n+r+1}$, we obtain
\newline
$1 +(\hat{a}_{r-1}+v)\theta(1)x^1+\cdot\cdot\cdot+(\hat{a}_1+v)\theta^{r-1}(1)x^{r-1}+1\theta^{r}(1)x^{r}\in\tilde{\mathcal C}.$ Then
\newline
$1+(\hat{a}_{r-1}+v)x^1+\cdot\cdot\cdot+(\hat{a}_1+v)x^{r-1}+x^{r}\in\tilde{\mathcal C}.$
By  Equation (~\ref{eq:hat}),  we obtain  $f_1^*(x)=1+ a_{r-1}x^1+\cdot\cdot\cdot+a_1x^{r-1}+x^{r}\in\tilde{\mathcal C}$, hence
$vf_1^*(x)=1+ v(a_{r-1}x^1+\cdot\cdot\cdot+a_1x^{r-1}+x^{r})\in\tilde{\mathcal C}$ by the Corollary  ~\ref{cor:struct1}, we have $vf_1^*(x)= vf_1(x)q(x)$, one necessary have $q(x)=1.$ Then $f_1^*(x)= f_1(x).$
\qed
\begin{thm}

\label{thm:reves3}
Let $\tilde{\mathcal C}=(f(x),g(x))$ be a skew cyclic code in $\tilde {R}_n$, where $f(x)$ is a polynomial of minimal degree  in $\tilde{\mathcal C}$ and is not a monic polynomial, $g(x)$ is a polynomial of least degree among the monic  polynomials in $\tilde{\mathcal C}$. If $\tilde{\mathcal C}$ is a reverse-complement code then  $f(x)$  and $g(x)$ are self-reciprocal.
\end{thm}
\pf
The proof is similar to the proof of the Theorem \ref{thm:reves1} and of Theorem \ref{thm:reves2}.
\qed

In the following, we provide sufficient  conditions for $\tilde{\mathcal C}$ being reverse-complement.
\begin{thm}
\label{thm:re1}
Let $\tilde{\mathcal C}=(f(x))$ be a skew cyclic codes in $\tilde {R}_n$, where $f(x)$ is monic polynomial of the degree minimal in $\tilde{\mathcal C}.$ If $v(x^n-1)/(x-1)\in\tilde{\mathcal C}$ and $f(x)$ is self-reciprocal then $\tilde{\mathcal C}$ is reverse-complement.
\end{thm}
\pf
Let $f(x)=1+a_1x+a_2x^2+\cdot\cdot\cdot+a_{r-1}x^{r-1}+x^r$ be a monic polynomial of the degree minimal in $\tilde{\mathcal C}$ and $c(x)\in\tilde{\mathcal C}$, we say that
$c(x)=q(x)f(x)$ where $q(x)\in R[x,\theta].$
\newline
$c(x)^*=(q(x)f(x))^*,$ by the Lemma ~\ref{lem:recip} we have $c(x)^*=q(x)^*f(x)^*$, since $f(x)$ is self-reciprocal then  $c(x)^*=q(x)^*f(x)\in\tilde{\mathcal C}$
for all $c(x)\in \tilde{\mathcal C}.$
We know that
\begin{equation}
\label{eq:nab1}
v+vx+\cdot\cdot\cdot+vx^{n-1}\in \tilde{\mathcal C}
\end{equation}
Now, let $c(x)=c_0+c_1x+c_2x^+\cdot\cdot\cdot+c_tx^t,$ we multiply the right polynomial $c(x)$ by $x^{n-t-1}$ we obtain
 $c(x)*x^{n-t-1} =c_0+c_1\theta(1)x+c_2\theta^{2}(1)x^2+\cdot\cdot\cdot+c_t\theta^{t}(1)x^t,$ then
\begin{equation}
\label{eq:nab2}
c(x)*x^{n-t-1}=c_0x^{n-t-1}+c_1x^{n-t}+\cdot\cdot\cdot+c_tx^{n-1}\in \tilde{\mathcal C}
\end{equation}
Combining (~\ref{eq:nab1}) and ( ~\ref{eq:nab2})  we obtain
\begin{equation}
\label{eq:nab3}
(v+vx+\cdot\cdot\cdot+vx^{n-t-2}+(c_0+v)x^{n-t-1}+\cdot\cdot\cdot+(c_t+v)x^{n-1})\in \tilde{\mathcal C},
\end{equation}
 leading to the following equality (using Equation ~\ref{eq:hat})
$$v+vx+\cdot\cdot\cdot+vx^{n-t-2}+\hat{c}_0x^{n-t-1}+\hat{c}_1x^{n-t}+\cdot\cdot\cdot+\hat{c}_{t-1}x^{n-2}+\hat{c}_tx^{n-1}=(c(x))^{rc}.$$
Therefore,  $(c^*(x)^{rc})^*=c(x)^{rc}\in \tilde{\mathcal C}.$
\qed
\begin{thm}
Let $C=(vf_1(x))$ be a skew cyclic code in  $\tilde {R}_n$, where $f_1(x)$ is a monic binary polynomial of lowest degree with $f_1(x)|(x^n-1)$. If $v(x^n-1)/(x-1)\in\tilde{\mathcal C}$ and $f_1(x)$ is self-reciprocal then $\tilde{\mathcal C}$ is reverse-complement.
\end{thm}
\pf
The proof is similar to the proof of Theorem \ref{thm:re1}.
\qed
\begin{thm}
Let $C=(f(x),g(x))$ be a skew cyclic codes in $\tilde {R}_n$, where $f(x)$ is a polynomial of degree minimal in $\tilde{\mathcal C}$ and is not monic polynomial, $g(x)$ is a polynomial of least degree among monic  polynomial in $\tilde{\mathcal C}$. If $v(x^n-1)/(x-1)\in\tilde{\mathcal C}$ and $f(x)$ and $g(x)$ are  self-reciprocal then $\tilde{\mathcal C}$ is reverse-complement.
\end{thm}
\pf
The proof is similar to the one of Theorem \ref{thm:re1}.
\qed
\subsection{Binary Image of Skew DNA Cyclic Codes}
We now recall the definition of the Gray map $\varphi$ for $\mathbb{F}_2+v\mathbb{F}_2$ to $\mathbb{F}_2,$ for each element of $\mathbb{F}_2$  expressed as  $a+vb$, where $a,b\in\mathbb{F}_2$. $\varphi(a+vb)=(a+b,a)$ i.e; $0\rightarrow(0,0)$, $1\rightarrow(1,1)$, $v+1\rightarrow(0,1)$, $v\rightarrow(1,0).$
\newline
According to the definition of the Gray map it is easy to verify that $\varphi$ is linear.

We can obtained the binary image of the DNA code by the map $\varphi$ and the map $\psi$, the DNA alphabet onto the set of the length 2 binary word given as following  $G\rightarrow(0,0)$, $A\rightarrow(1,1)$, $T\rightarrow(0,1)$, $C\rightarrow(1,0).$
We have the following property of the binary image of a DNA skew cyclic code.
\begin{cor}
The map $R\mapsto{\mathbb{F}_2}^n$ is distance preserving linear isometry, thus if $\tilde{\mathcal C}$ is a skew cyclic DNA code over $R$ then $\varphi(\tilde{\mathcal C})$ is a skew quasi-cyclic DNA code of the length $2n$ and of the index $2$.
\end{cor}
\pf
The proof is similar to the proof of Lemma \ref{lem:image}
\qed
\begin{table}[h]
\caption{$[10,6,2]$ reverse-complement Skew Cyclic DNA Code }
\begin{center}
\begin{tabular}{|l|l|l|l|}
  \hline
  GGGGGGGGGG&CCCCCCCCCC&CCCCCGGGGG&GGGGGCCCCC\\
  GGGGCCCCCG&CCCCGGGGGC&CCCCGCCCCG&GGGGCGGGGC\\
  GGGTTTTTGG&CCCAAAAACC&CCCATAAACG&GGGTATTTGC\\
  GGGTAAAACG&CCCATTTTGC&CCGGGCCGGG&GGCCCGGCCC\\
  GGCCCCCGGG&CCGGGGGCCC&CCGGCGGCCG&GGCCGCCGGC\\
  GGCCGGGCCG&CCGGCCCGGC&CCGTATTACG&GGCATAATGC\\
  GGCAAAATCG&CCGTTTTAGC&CCGTTAATTG&GGCAATTAAC\\
  GGCAAAATGG&CCGTTTTACC&CATTAACGGG&GTAATTGCCC\\
  GTAAAACGGG&CATTTTGCCC&CAGTATGCCG&GTCATACGGC\\
  GTAACCATTG&CATTGGTAAC&CAAGCGTACG&GTTCGCATGC\\
  GTACGGTACG&CATGCCATGC&CAATTACCGG&GTTAATGGCC\\
  GTACCCATGG&CTAGGGTACC&CAAAATGGGG&GTTTTACCCC\\
  GATTTTGGGG&CTAAAACCCC&CAAATACCCG&GTTTATGGGC\\
  GTTTAACCCG&CAAATTGGGC&CAACGCAACG&GTTGCGTTGC\\
  GTTGCCAACG&CAACGGTTGC&CAACCGTTGG&GTTGGCAACC\\
  GTTGGGTTGG&CAACCCAACC&CCACGCAACG&GGTGCGTTGC\\
  \hline
\end{tabular}
\end{center}
\end{table}

\end{document}